\documentclass[preprint,showpacs,preprintnumbers,amsmath,amssymb]{revtex4}

\usepackage{graphicx}
\usepackage{dcolumn}
\usepackage{bm}

\begin{document}


\title{Fragmentation of  1.2 A GeV $^7$Be nuclei in nuclear photographic emulsion }

\author{N.~K.~Kornegrutsa}
   \affiliation{Joint Insitute for Nuclear Research, Dubna, Russia}
 \author{D.~A.~Artemenkov}
   \affiliation{Joint Insitute for Nuclear Research, Dubna, Russia} 
 \author{V.~Bradnova}
   \affiliation{Joint Insitute for Nuclear Research, Dubna, Russia}   
 \author{P.~I.~Zarubin}
     \email{zarubin@lhe.jinr.ru}    
     \homepage{http://becquerel.jinr.ru}
   \affiliation{Joint Insitute for Nuclear Research, Dubna, Russia} 
 \author{I.~G.~Zarubina}
   \affiliation{Joint Insitute for Nuclear Research, Dubna, Russia}   
  \author{R.~R.~Kattabekov}
   \affiliation{Institute for Physics and Technology, Uzbek Academy of Sciences, Tashkent, Republic of Uzbekistan} 
 \author{K.~Z.~Mamatkulov}
   \affiliation{Djizak State Pedagogical Institute, Djizak, Republic of Uzbekistan}  
 \author{P.~A.~Rukoyatkin}
   \affiliation{Joint Insitute for Nuclear Research, Dubna, Russia} 
 \author{V.~V.~Rusakova}
   \affiliation{Joint Insitute for Nuclear Research, Dubna, Russia}     
   
\date{\today}

\begin{abstract}
\indent  The charge topology of peripheral fragmentation of 1.2 A GeV $^7$Be nuclei in a nuclear emulsion is presented.
 The dissociation of $^7$Be nuclei via the channels $^7$Be$\rightarrow ^4$He + $^3$He, $^7$Be$\rightarrow$2$^3$He + n and
 $^7$Be$\rightarrow ^4$He + 2$^1$H is considered in detail. It is found that in the channel $^7$Be$\rightarrow ^4$He + 2$^1$H,
 events related to the channel $^7$Be$\rightarrow ^6$Be + n with the cascade decay $^6$Be$\rightarrow ^4$He + 2p account for
 about 27\%.

\end{abstract}

\maketitle

\indent Due to the possibility of complete observation of the charged component of the fragmentation products, layers of nuclear emulsion longitudinally
 exposed to a beam of light relativistic nuclei provide wide possibilities for studying the cluster structure of light neutron-deficient nuclei \cite{ref1},
 \cite{ref2}, \cite{ref3}. The present work on the study of the dissociation of $^7$Be nuclei is a continuation of a series of studies conducted by the
 BECQUEREL collaboration \cite{ref1} on the cluster structure of light nuclei \cite{ref1} - \cite{ref9}. The $^7$Be nucleus is of interest as a source of
 information on the configurations $^3$He + $^4$He, $^3$He + $^3$He + n, $^6$Li + p, $^6$Be + n. In addition, information on the fragmentation of this
 nucleus is important for understanding the cluster structure of subsequent nuclei at the boundary of the proton bond $^7$Be, $^9$C and $^{12}$N, since the
 $^7$Be nucleus plays the role of the core in them.\par 
\indent A nuclear photoemulsion was exposed to a mixed beam of $^7$Be, $^{10}$C, and $^{12}$N nuclei formed by separating the products of charge exchange
 and fragmentation of primary $^{12}$C nuclei with energy of 1.2 A GeV at the JINR Nuclotron \cite{ref6}. The search for events was carried out using
 primary tracks without sampling. The charge of the beam nuclei and relativistic fragments was identified by visually counting the density of $\delta$-electrons.
 As a result, the average range before the interaction of $^7$Be nuclei with photoemulsion nuclei in this work is 14.2$\pm$0.2 cm. In \cite{ref4}, the
 range of $^7$Be nuclei for a photoemulsion of the same type is 14.0$\pm$0.8 cm.\par
 \indent Viewing the exposed emulsions and subsequent classification of the tracks made it possible to obtain a picture of the charge topology of the
 peripheral fragmentation of the $^7$Be nucleus. Table 1 shows the distribution of 289 N$_{ws}$ events by fragmentation channels that were not accompanied
 by target fragments (\lq\lq white\rq\rq stars) and mainly related to interactions on the Ag and Br emulsion nuclei. For comparison, the distribution of
 380 $^7$Be N$_{tf}$ fragmentation events accompanied by tracks of target fragments is shown. It is noteworthy that a significant proportion of the found
 events (about 90\%) fall on the $^7$Be$\rightarrow$2He and $^7$Be$\rightarrow$He + 2H channels, corresponding to thresholds of 1.6 MeV and 9.3 MeV.
 The $^7$Be$\rightarrow$4H channel with a high energy formation threshold (37.6 MeV) corresponds to a lower probability.\par
 
\begin{table}[t]
\caption{Distribution of $^7$Be nuclei dissociation channels for \lq\lq white\rq\rq stars N$_{ws}$ and events with target 
fragments or produced mesons N$_{tf}$.}
\centering
\label{tab:1}       
\begin{tabular}{lcccc}
\hline\noalign{\smallskip}
Channel & 2He & He + 2H & 4H & Li + H\\[3pt]
\hline\noalign{\smallskip}
N$_{ws}$ & 115 & 157 & 14 & 3 \\
N$_{tf}$ & 154 & 226 & - & -\\
\hline\noalign{\smallskip}
\end{tabular}
\end{table}

\indent Identification of relativistic He and He fragments by multiple scattering became one of the main tasks of the study. To automate the identification
 process, a classifier program based on a neural network was developed. The results of modeling in Geant4 of the passage of $^7$Be nuclei with an energy of
 1.2 A GeV in an emulsion were used as a training sample. Table 2 shows the distribution of events by the $^7$Be$\rightarrow$2He channels based on the
 results of the classification of He fragments. Based on the statistics of 174 events for which all angular measurements were carried out, it was possible
 to completely identify only 79 events. Since the identification was carried out without any sample, Table 2 gives an idea of the ratio of the cluster
 configurations $^3$He + $^4$He and 2$^3$He + n in the structure of the $^7$Be nucleus. The channel $^7$Be$\rightarrow ^3$He + $^4$He dominates over the
 channel $^7$Be$\rightarrow$2$^3$He, which indicates a higher probability of a two-cluster configuration in the structure of $^7$Be, compared to the
 three-particle 2$^3$He + n. At the same time, the probability of 2$^3$He + n is significant and amounts to about 30\%, which is consistent with previously
 obtained data \cite{ref4}.\par

\begin{table}[t]
\caption{Distribution of identified  N$_{ws}$ and N$_{tf}$ events in $^7$Be$\rightarrow$2He fragmentation channels.}
\centering
\label{tab:2}       
\begin{tabular}{lcc}
\hline\noalign{\smallskip}
Channel & $^3$He + $^4$He &   $^3$He + $^3$He\\[3pt]
\hline\noalign{\smallskip}
N$_{ws}$ & 32 & 14 \\
N$_{tf}$ & 24 & 9 \\
\hline\noalign{\smallskip}
\end{tabular}
\end{table}

\indent Figure 1 shows the distribution of the polar angle of emission $\theta$ of the He fragments of the entire group of measured events, as well as
 $^3$He and $^4$He for completely identified events  - dotted and hatched histograms, respectively. The parameters of the Rayleigh distributions describing
 the spectrum of angles $\theta$ for $^3$He and $^4$He are $\sigma_{\theta}$($^3$He) = (17$\pm$2)$\times$10$^{-3}$ rad and $\sigma_{\theta}$($^4$He) = (16$\pm$2)
 $\times$10$^{-3}$ rad. Estimates for these parameters using the statistical model \cite{ref10}, \cite{ref11} were $\sigma_{\theta}$($^3$He) = 20$\times$10$^{-3}$
 rad and $\sigma_{\theta}$($^4$He) = 15$\times$10$^{-3}$ rad. The parameters of the Rayleigh distributions describing the spectrum of transverse momentum
 P$_{T}$ in the approximation of conservation of momentum per nucleon of the parent nucleus \cite{ref2}, \cite{ref3} for the fragments $^3$He, $^4$He
 (Fig. 2) are equal to $\sigma_{PT}$($^3$He) = (97$\pm$7) MeV/c and  $\sigma_{PT}$($^4$He) = (125$\pm$17) MeV/c, respectively. The value for $^4$He is in
 good agreement with the value of the statistical model $\sigma_{PT}$($^4$He) = 121 MeV/c.\par
\indent The distribution of events of the channels $^7$Be$\rightarrow$2$^3$He and $^7$Be$\rightarrow ^3$He+$^4$He by excitation energy Q (Q = M$^*$ - M) of
 the fragment system defined as the difference between the invariant mass of the fragmenting system M$^*$ and the total mass of the fragments M, is shown
 in Fig. 3. The invariant mass of the fragment system M$^*$ is determined according to the expression M$^{*2}$ = $(\Sigma P_{j})^2$ = $\Sigma(P_{i}P_{k})$,
 where P$_{i}$, P$_{k}$ are the 4-momenta of the fragments in the approximation of conservation of momentum per nucleon of the parent nucleus.
 The obtained values of Q$_{2He}$ for events of the channel  $^7$Be$\rightarrow ^3$He+$^4$He  are located in the region of excitation levels of the $^7$Be nucleus.\par
 \indent One of the objectives of this experiment was to detect  $^7$Be$\rightarrow$2$^3$He events characterized by Q$_{^3He^3He}$ values lying in the
 range of 100-200 keV, similar to those observed in \cite{ref7}. The resulting spectrum contains a group of 4 events for which Q$_{^3He^3He}$ values are
 located in the range from 200 to 400 keV (Fig. 3, dotted histogram in the inset). These data do not exclude the possible existence of a resonance state
 of 2$^3$He, discussed in \cite{ref7}.\par
 \indent For the $^9$Be and $^{10,12}$C nuclei, a significant contribution from cascade fragmentation to form an unstable $^8$Be nucleus is established \cite{ref2},
 \cite{ref3}, \cite{ref8}, \cite{ref9}. In the case of the $^7$Be isotope, there is a possibility of cascade fragmentation of $^7$Be to form the unstable $^6$Be with a
 threshold of 1.37 MeV over $^4$He + 2p. Figure 4 shows the distribution of 130 measured events of the $^7$Be$\rightarrow ^4$He + 2p fragmentation channel by the differences
 in the invariant mass of the resulting fragments of the alpha particle, two protons, and the sum of their masses Q$_{^4He+2p}$. The region Q$_{^4He+2p} <$ 6 MeV 
 (Fig. 4, histogram in the inset) indicates the presence of a significant proportion ($\approx$27\%) of $^7$Be$\rightarrow ^6Be\rightarrow ^4$He + 2p events. A feature of
 this group of events is a \lq\lq narrower\rq\rq distribution by the value of the total transverse momentum P$_{Tsum}$($^4$He + 2p) (Fig. 5) compared to the distribution
 for the entire sample. The Rayleigh distribution parameter is $\sigma_{PT}$ = 124$\pm$20 MeV/c, which is greater than the value calculated by the statistical model
 $\sigma_{PT}$= 86 MeV/c for $^6$Be. This difference may be due to the fact that the statistical model does not fully take into account the reaction mechanism.\par
 \indent The question of the contribution of the $\alpha$ + p resonance decay of $^5$Li with an energy of 1.69 MeV and a width of 1.5 MeV is of independent importance,
 since the threshold for the formation of the $^5$Li + p system is 0.35 MeV higher than the ground state of $^6$Be. Despite the absence of a clear signal due to combinatorial
 complexity, the Q$_{\alpha p}$ spectrum (Fig. 6) does not contradict the possible contribution of the $\alpha$ + p resonance decays of $^5$Li.\par
 \indent In conclusion, we list the main results of this study. For the first time, a detailed study of the fragmentation of $^7$Be nuclei on photoemulsion nuclei was carried out.
 Angular and momentum spectra of the resulting fragments were obtained. The most probable modes in peripheral fragmentation are events accompanied by the formation of 2He and
 He + 2H. For events with the formation of 2He, a distribution between the channels $^3$He + $^4$He and 2$^3$He in a ratio of $\approx$70\% and $\approx$30\% is typical.
 The problem of observing the resonance state of 2$^3$He in the dissociation of $^7$Be requires further increase in statistics; an indication of its presence was obtained in
 the case of the $^9$C nucleus \cite{ref7}. Analysis of the distribution by the excitation energy value Q$_{^4He+2p}$ indicates the presence of a contribution of about 27\%
 of events from the channel  $^7$Be$\rightarrow ^4$He + 2p accompanied by a chain of transformations $^7$Be$\rightarrow ^6Be\rightarrow ^4$He + 2p.\par
\indent  The authors express their gratitude to S.P. Kharlamov for discussion of the results. This work was supported by the grant of the Russian Foundation for
 Basic Research 12-02-00067, as well as grants from the plenipotentiary representatives of Bulgaria and Romania at JINR.\par

\begin{figure}
\centering
  \includegraphics[width=5in]{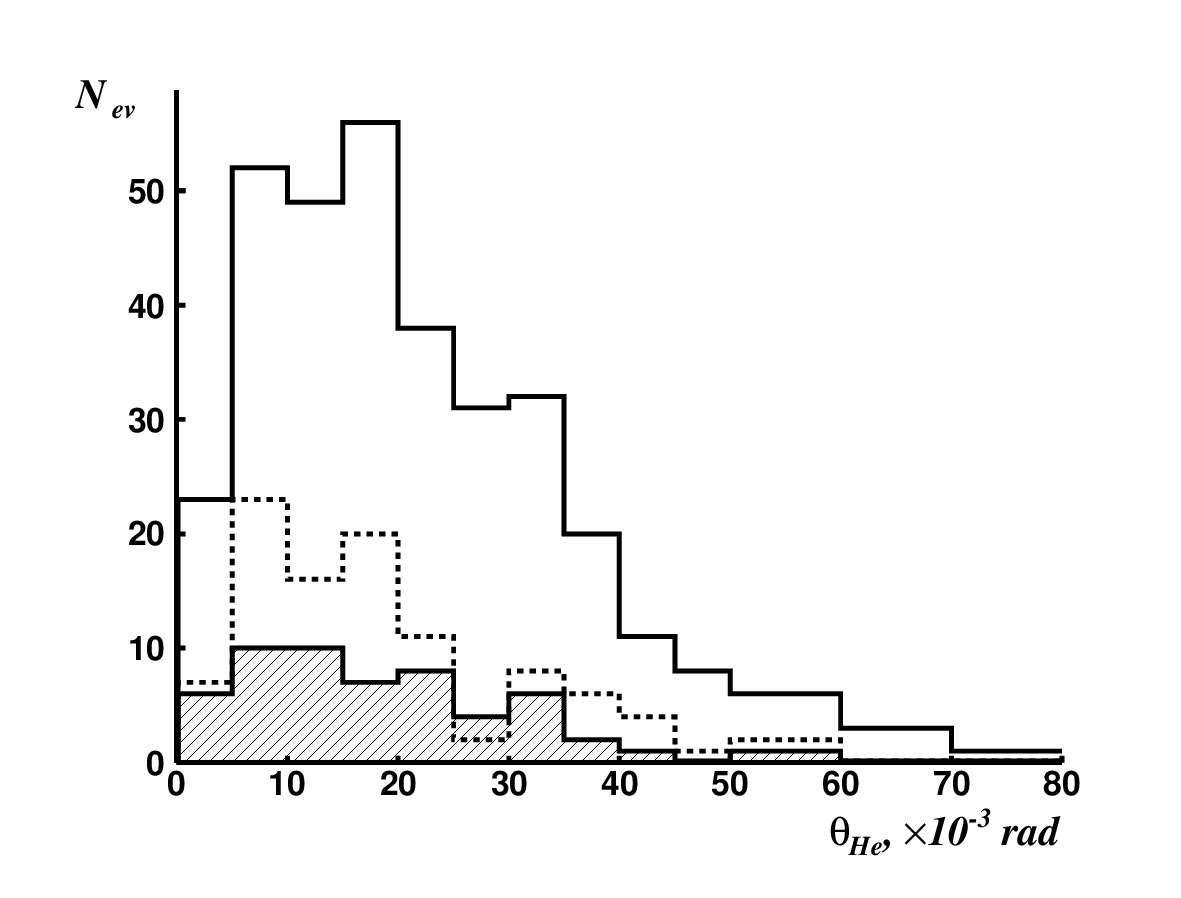}
\caption{\label{fig:1}  Distribution of He fragments of  $^7$Be$\rightarrow$2He channel over polar angle of emission $\theta$ for entire sample of measured events -- solid
 line for fully identified,$^3$He -- dashed, $^4$He -- hatched histograms.}
\end{figure}
\begin{figure}
\centering
  \includegraphics[width=5in]{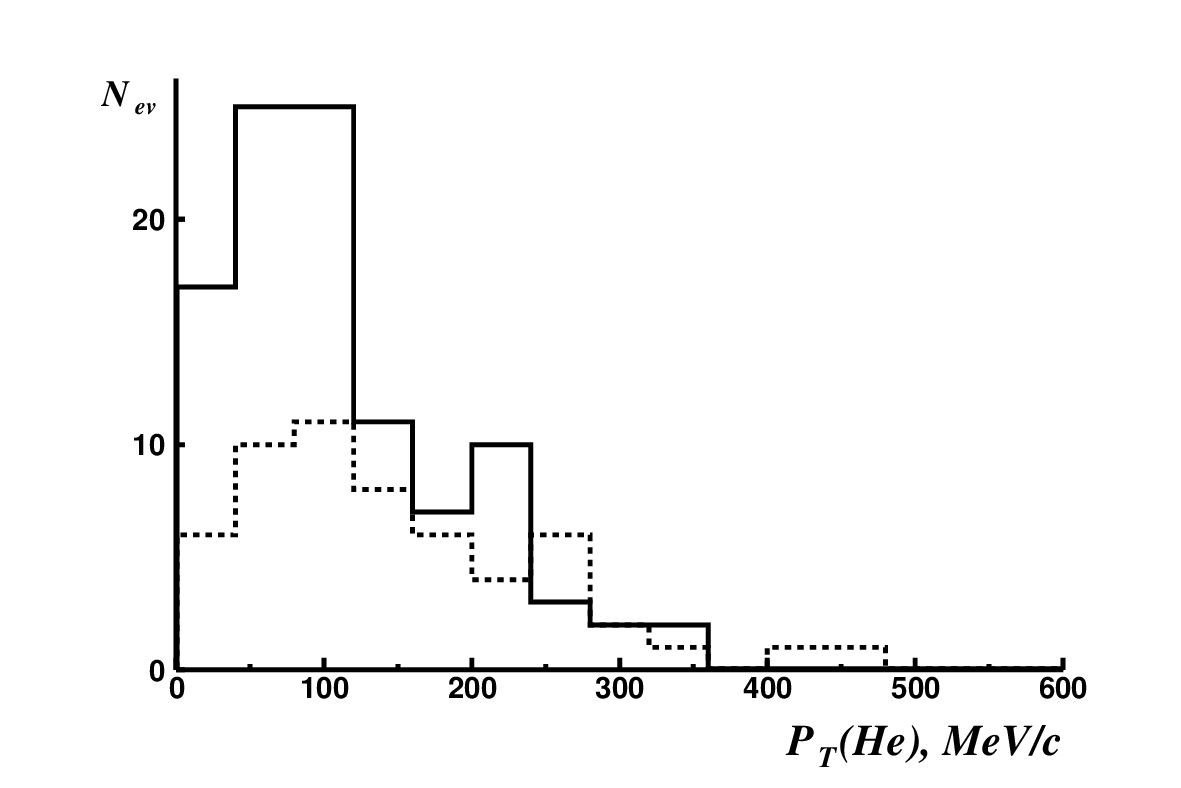}
\caption{\label{fig:2} Distribution of identified 3,4He fragments of $^7$Be$\rightarrow$2He channel over value of transverse momentum P$_{T}$ 
($^3$He -- solid, $^4$He -- dashed histograms)}
\end{figure}

\begin{figure}
\centering
  \includegraphics[width=5in]{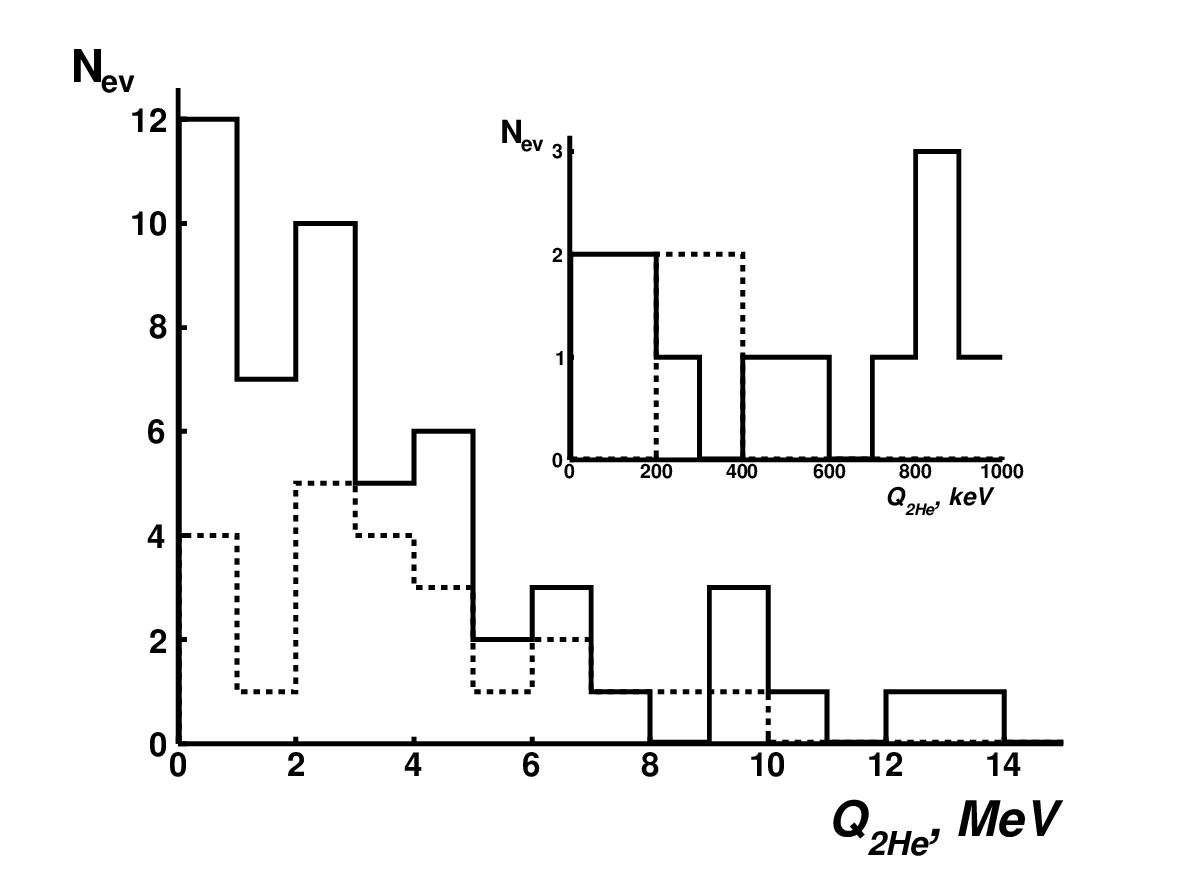}
\caption{\label{fig:3} Distribution of events of channels  $^7$Be$\rightarrow ^3He+^4$He and 2$^3$He by excitation energy Q
 (solid and dotted lines of the histograms, respectively). The inset shows the histograms for values of Q $<$ 1 MeV.}
\end{figure}
 
 \begin{figure}
\centering
  \includegraphics[width=5in]{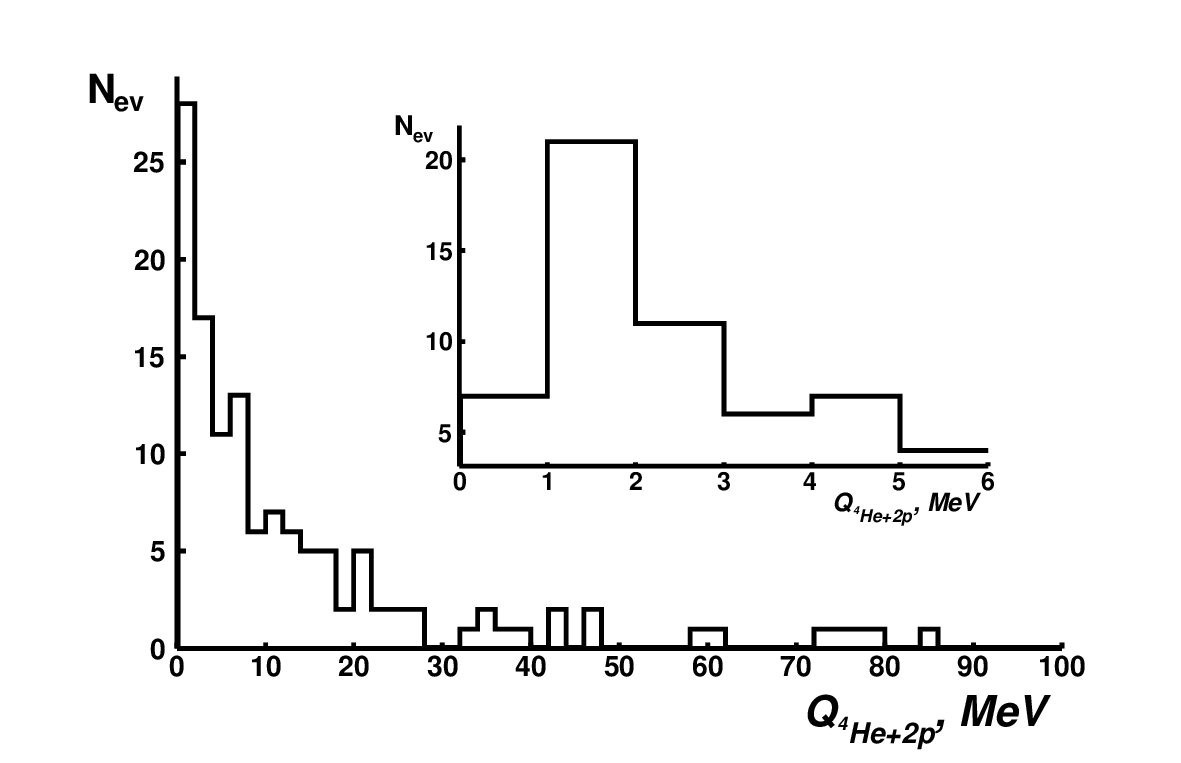}
\caption{\label{fig:4} Distribution of events of channel $^7$Be$\rightarrow ^4$He + 2p over excitation energy value Q$_{^4He+2p}$.}
\end{figure}

\begin{figure}
\centering
  \includegraphics[width=5in]{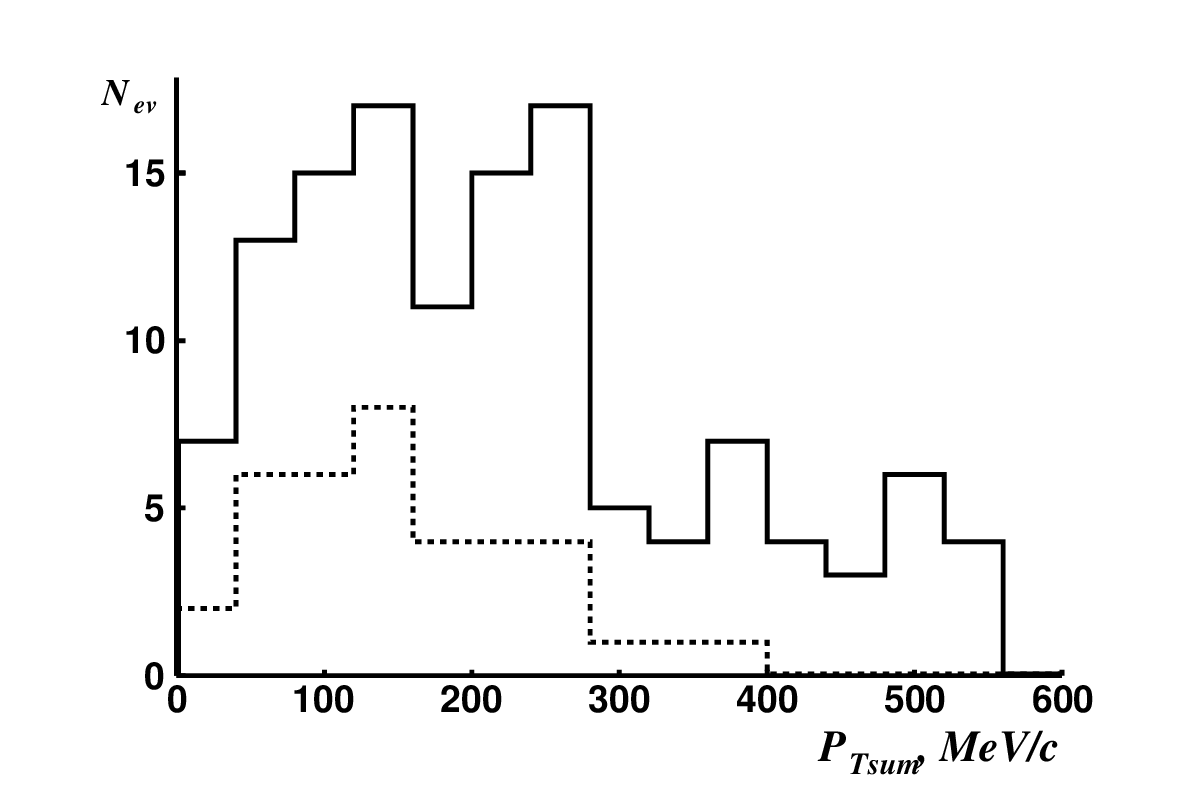}
\caption{\label{fig:5} Distribution of events of $^7Be\rightarrow ^{4}$He + 2p channel over value of total transverse momentum of fragments P$_{Tsum}$($^4$He + 2p);
 dotted histogram corresponds to $^7Be\rightarrow ^{6}Be\rightarrow ^{4}$He + 2p channel.}
\end{figure}
 
  \begin{figure}
\centering
  \includegraphics[width=5in]{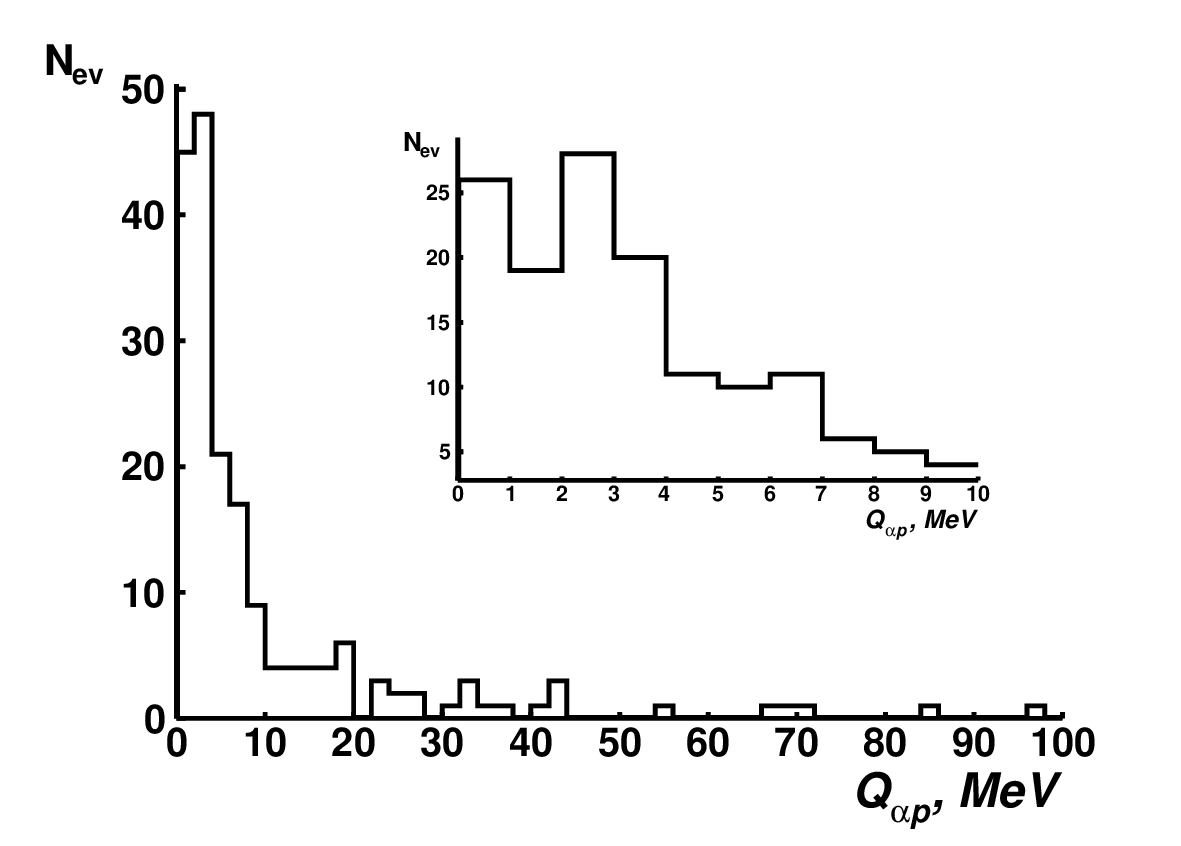}
\caption{\label{fig:6} Distribution of events of $^7$Be$\rightarrow ^4$He + p channel over value of excitation energy Q$_{^4He+p}$
 (events attributed to $^6$Be fragmentation are excluded from the histogram).}
\end{figure}


\newpage


\begin{thebibliography}{}
\bibitem{ref1} 
 The BECQUEREL Project, http://becquerel.jinr.ru/
\bibitem{ref2}
 Belaga~V.~V. et al., \lq\lq Coherent Dissociation $^{12}$C$\rightarrow3\alpha$ in Lead-Enriched Emulsion at 4.5 GeV/c per Nucleon\rq\rq
 Phys. Atom. Nucl. \textbf{58}, 1905 (1995); arXiv:1109.0817. 
\bibitem{ref3}
 Artemenkov~D.~A. et al., \lq\lq Features of the $^9$Be$\rightarrow$2He fragmentation in an emulsion for an energy of 1.2 GeV per nucleon\rq\rq
 Phys. Atom. Nucl. \textbf{70}, 1222(2007); nucl-ex/0605018. https://doi.org/10.1134/S1063778807070125
\bibitem{ref4}
Peresadko~N.~G. et al., \lq\lq Fragmentation channels of relativistic $^7$Be nuclei in peripheral interactions\rq\rq 
Phys. Atom. Nucl. \textbf{70}, 1266 (2007); nucl-ex/0605014. https://doi.org/10.1134/S1063778807070137
\bibitem{ref5} 
Stanoeva~R. et al., \lq\lq Electromagnetic dissociation of relativistic $^8$B nuclei in nuclear track emulsion\rq\rq 
Phys. Atom. Nucl. \textbf{72}, 690 (2009); arXiv: 0906.4220. https://doi.org//10.1134/S1063778809040140
\bibitem{ref6}
Kattabekov~R.~R., Mamatkulov~K.~Z. et al., \lq\lq Exposure of nuclear track emulsion to a mixed beam of relativistic $^{12}$N, $^{10}$C, and $^7$Be nuclei\rq\rq
 Phys. Atom. Nucl. \textbf{73}, 2110 (2010); arXiv:1104.5320. https://doi.org/10.1134/S1063778810120161
\bibitem{ref7} 
Krivenkov~D.~O.  et al., \lq\lq Coherent dissociation of relativistic $^9$C nuclei\rq\rq 
Phys. Atom. Nucl. \textbf{73}, 2103 (2010); arXiv:1104.2439. https://doi.org/10.1134/S106377881012015X
\bibitem{ref8} 
 Artemenkov~D.~A. et al., \lq\lq Dissociation of Relativistic $^{10}$C Nuclei in Nuclear Track Emulsion\rq\rq
 Few Body Syst. \textbf{50}, 259 (2011); arXiv:1105.2374. https://doi.org/10.1007/s00601-011-0223-z
\bibitem{ref9} 
Artemenkov~D.~A. at al., \lq\lq Dissociation of Relativistic $^{10}$C Nuclei in Nuclear Track Emulsion\rq\rq
 Int. J. Mod. Phys. E \textbf{20}, 993 (2011); arXiv: 1106.1749. https://doi.org/10.1142/S021830131101912X
\bibitem{ref10} 
Feshbach~H. and Huang~K. \lq\lq Fragmentation of relativistic heavy ions\rq\rq
 Phys. Lett. 47B, 300 (1973). https://doi.org/10.1016/0370-2693(73)90607-2
\bibitem{ref11}
Goldhaber~A.~S.  \lq\lq Statistical models of fragmentation processes\rq\rq
Phys. Lett. 53B, 306 (1974). https://doi.org/10.1016/0370-2693(74)90388-8
 
\end{thebibliography}
\end{document}